\documentstyle{article}
\begin{document}
\begin{center}
{\bf Applying the DFT representation for describing the order-chaos transition in H\'enon mapping dynamics}\\
\vspace{1cm}
{\it Carlos R. Fadragas(1), Ruben Orozco-Morales(2), Juan V. Lorenzo-Ginori(2),\\(1)Physics Department, Central University of Las Villas, Santa Clara, Cuba,\\(2)Centre for Studying Electronics and Information Technologies,\\ Central University of Las Villas, Santa Clara, Cuba.}
\end{center}
\vspace{1cm}
\begin{center}
{\bf Abstract}
\end{center}
The DFT representation is applied, to a ordered-chaotic finite-duration sequences set generated by the H\'enon mapping, for describing the order-chaos transition. This representation has the advantage that it may be applied to a relatively shorter discret-time series. The bifurcation diagram is produced and the largest Lyapounov exponent is calculated for each time series of the set, showing a good agreement with the results obtained by the DFT representation. The threshold value of the control parameter for the order-chaos transition was determined to be A=1.0532.
\section{\bf Introduction}

The Discrete Fourier Transform (DFT) representation method has been yet applied for describing the order-chaos transition in a dynamical model represented by the Poincar\'e mapping \cite{Fadragas}. This representation has the advantage that can operate over a relatively shorter discrete-time series, and it was demonstrated that the DFT representation of a chaotic discrete-time series is possible, nevertheless this discrete-time series type is to be a nonlinear one. Literature reports that transition from a regular to a chaotic bahavior has been studied in many dynamical system, including bilogical systems \cite{Bleek,Khadra}.

In this work, the H\'enon mapping dynamics is considered, and the DFT representation is applied to describing the order-chaos transition. The H\'enon mapping is a dynamical model widely used that can produce a chaotic discrete-time series, and therefore it is important to evaluate the largest Lyapounov exponent for each of the discrete-time series produced.

In 1968, Michel H\'enon \cite{Henon} proposed a quadratic mapping of the plane as a way for stuying a dynamical system such as the motion of a asteroid or a charged particle in an accelerator. H\'enon mapping is based on the idea that one can reduce the study of a conservative system with two degree of freedom to the study of area-preserving mappings of the plane. H\'enon mappings simulate the behavior of a physical system and show that many such systems are more complex than previously imaged. The nature of this complexity just begins to be understood now a day by researchers.

Wolf et al \cite{Wolf} included the H\'enon mapping \cite{Henon}in their study for determining Lyapounov exponents from a discrete-time series. The Lyapounov exponents spectrum was calculated to be $\lambda_1=0.603$ bit/iter, and  $\lambda_2=-2.34$ bit/iter. The Lyapounov dimension was found to be $D_I=1.26$.

A control technique has been developed by Ott et al \cite{Ott}. It is based on the fact that there are an infinite number of unstable periodic orbits embedded within a chaotic attractor. This method is a powerful experimental tool that models the dynamics of a system directly from its discrete-time series. The Ott-Grebogi-Yorke technique has been used to control many different chaotic system. The principles of this technique have been used by an another technique called ¨tracking¨ to stabilize underlying unstable periodic orbits in nonchaotic systems. To investigate the feasibility of using this method, Christini and Collins \cite{Christini} considered the H\'enon mapping with additive noise, as given by the expression $x[n]=1.0-A*x[n-1]^2+B*x[n-2]+\xi[n]$, where A is an adjustable parameter, B=0.3, and $\xi[n]$, is a Gaussian white noise with zero mean and standard deviation known. They used A as a control parameter, varying into the interval [0.2,1.2], to produce a bifurcation diagram of the noise-free model. It was their interest to kepp the H\'enon mapping out of the chaotic region by setting A to 1.00. At this condition, the H\'enoin mapping exhibits a stable period-4 cycle.

Khadra et al \cite{Khadra} applied a statistical approach to chaos identification in discrete-time series. The method was applied to numerical data generated by chaotic systems and to heart rate variability signals of normal subjets and heart transplant recipients. To demonstrate the power of this technique they applied it to random and chaotic data. The random data were generated from an AR random process defined by $x[n+1]=0.9*x[n]+y[n]$, where y[n] are white Gaussian random numbers with zero mean and unit variance. The chaotic data were generated from the H\'enon mapping defined as above, using A=1.4, and B=0.3. 

Christini et al \cite{Christini} developed a real-time, model-independent control technique for chaotic and nonchaotic systems. For testing this technique they considered the H\'enon mapping with an adjustable parameter $A_n$, as given by $x[n]=1.0-A_n*x[n-1]^2+B*x[n-2]$, where B=0.3.

In a recent work of Iasemidis et al \cite{Iasemidis} about epilepsy, the voltages of the recorded signal x(t) at several points in time $[t, t+\tau,\cdots, t+(D_{e}-1)*\tau]$, separated by a given time delay $\tau$ were converted to a $D_{e}-dimensional$ vector $X(t)$. The process was repeated by increasing $t$ to obtain successive points in the phase space, creating a $D_{e}-dimensional$ structure. The phase space portrait can be analyzed for determining the presence of an attractor and its dimension. The presence of an attractor of low dimension means that of the many modes of the system only a portion of them are active during that period of observation. Thus, the effective phase space portrait dimension is much smaller than the full dimension of the equations governing the dynamics of the system. They found that the epileptogenic focus generates signals characteristic of a nonlinear dynamical system, and that analysis of signals from the centre of the epileptogenic focus demonstrated the presence of strange attractor of low dimensionality and with positive value for largest Lyapounov exponent (LLE) during the inmediate preictal period, during the seizure, and following the postictal period. They used the phase-space portrait method for each stage of the seizure process, and calculated for each stage the LLE of the signal.

The electrocorticogram (ECoG) is a highly nonstationary signal. This condition must be taken into account by the authors in order to modify the Wolf algorithm of the LLE estimating. In this frame, The H\'enon mapping was used for testing the LLE estimating algorithm. They considered the H\'enon mapping as given by 
\begin{eqnarray}
x[n+1]=1.0-A*x[n]^2+B*y[n],\\
y[n+1]=B*x[n]
\end{eqnarray}

with A=1.4, B=0.3, and the initial condition to be $(x[1],y[1])=(0.631,0.189)$. They chosen N=2000 data points, with $D_{e}$=2, and $\tau$=1 point. The LLE estimated value was 0.577 bit/iter, that is close to that obtained by Wolf et al, that is, 0.603 bit/iter.

Dynamical control of excitable biological systems is often complicated by the difficult and unreliable task of precontrol identification of unstable periodic orbits. However, chaos control techniques have been applied to a number of excitable biological systems. One particular control technique, proportional perturbation feedback control, uses isolated electrical stimuli to cause the cells to fire at a specified time, thus directly altering the variable of interest, in interexcitation interval. Christini et al \cite{Christini} illustrated the stable manifold placement characterization and tracking technique using the chaotic H\'enon mapping defined as given by $x[n+1]=A+B*x[n-1]-x[n]^2$, where A=1.4 and B=0.3, and x[n] representing the nth interexcitation interval. Here, the chaotic H\'enon mapping was chosen because its uncontrolled dynamics have no stable periodic orbit and because its unstable flip saddle unstable periodic orbit is of the same type as that reported experimentally in studies of excitable media spike trains.

Literature reports many other studies that include the H\'enon mapping model to produce a chaotic time series. Generally, the control parameter value chosen for producing a chaotic discrete-time series is A=1.4, but another value that satisfies the condition 1.0532$\leq$ A$\leq$1.4 may be used, for which all them the corresponding LLE is positive.

The aim of this research is to evaluate the ability of DFT representation of a discrete-time series for describing the order-chaos transition in the H\'enon mapping dynamics, where the control parameter A can be adjusted in order to obtain a ordered or a chaotic regime as one decides, and comparing this method with the LLE method.

In this work, a 120 discrete-time series set, with 1024 data points in each discrete-time series, was produced. For each discrete-time series, its 1024 DFT representation was obtained. By observating either the amplitude spectrum and the phase spectrum behaviors with the control parameter A of the system, a well defined value of this control parameter may be determined, beyond which the behaviour either the amplitude spectrum and the phase spectrum becomes quite irregularly, in contrast with the regular behavior of these magnitudes in the region before the indicated value above. There exists a complete correspondence when this result is compared with that obtained from the bifurcation diagram and from the LLE plot, both as a function of the control parameter A. The threshold value for order-chaos transition for the control parameter A is 1.0532, which corresponding to a LLE value close to zero.
\section{\bf Methods and Materials}
\subsection{\bf Obtention of the discrete-time series set}

A H\'enon mapping \cite{Henon} is an area-preserving map of the plane given by
\begin{eqnarray}
x[n+1]=x[n]*Cos\phi-(y[n]-x[n]^2)*Sin\phi,\\
x[n+1]=x[n]*Sin\phi+(y[n]-x[n]^2)*Coos\phi
\end{eqnarray}

Here $\phi$ is the phase angle, chosen as a fixed constant on the interval [0,$\pi$]. An initial point (x[1],y[1]) is chosen each times to generate one orbit of the mapping by applying a iterative process, and modifying the initial point until a typical mapping containing 15 or 20 orbits is produced. It is important to choose an adecuate starting point for generating the orbit.

Using the equations above, a typical-38 orbit H\'enon mapping was obtained. For generating orbits, two linear space for the starting point was created, that is, 
\begin{eqnarray}
x[1]=linspace(0.098,0.718,38)\\
y[1]=linspace(0.061,0.581,38)
\end{eqnarray}

The phase angle $\phi$ was chosen as $\phi$=1.111 rad. A starting point is chosen each times to generate a 1024 point-orbit of the mapping. 

The H\'enon mapping can be too formulated as given by
\begin{eqnarray}
x[n+1]=1.0-A*x[n]^2+B*y[n],\\
y[n+1]=B*x[n]
\end{eqnarray}

where B=0.3, and A is a adjustable control parameter. With the initial condition (x[1],y[1]) specified, a discrete-time series with a given length by a recursive action is obtained, each times. The nature of the behavior of a data sequence obtained can be modified by chosen conveniently a value of the control parameter A. The interval [0.2,1.4] was chosen, and a set of evenly spaced values of A with a step $\delta$=0.01 was produced to obtain a family of 120 discrete-time series, each of them containing $N=2^{10}$ samples. The building up of the histogram for each discrete-time series allow us to have inmediatly a simple statistical characterization of the discrete-time series.

In order to make much easier further computations, data points were organized into a rectangular matrix containing 120 columns, being each of them a time series with the length $N=1024$ data points. 
\subsection{\bf Determining of the largest Lyapounov exponent $\lambda_1$}

A system containing one or more positive exponents is to be defined as chaotic. For calculating the Lyapounov exponents stectrum some algorithms have been developed. One of the most used algorithm that reported by Wolf et al \cite{Wolf}. From the exponents spectrum, the largest exponent, $\lambda_{1}$, decides the behaviour of the dynamical system. For calculating the largest exponent, one can too refer the algorithm proposed by Rosenstein et al \cite{Rosenstein}. Details of these algorithm can be found in refered papers.  

The largest Lyapounov exponent,$\lambda_1$, for each time series in the set was determining using a professional software \cite{Sprott}. The selection of the time delay, $\tau$, and the embedding dimension, $D_{e}$, required for reconstructing  the phase space is part of the problem. The value $\tau$=1 was chosen considering the optimal filling of the phase space method\cite{Buzug}. Figure 4 depicts the effect of optimal filling of the phase space of the attractor for the value of $\tau$ indicated above. For embedding dimension the value $D_{e}$=3 was chosen.
\subsection{\bf Application of DFT representation: Amplitude spectrum and phase spectrum}

The DFT representation of a finite-length discrete-time series $x[n];n=0,1,2,\ldots,N-1$, is given by
\begin{eqnarray}
x[n]=\frac{1}{N}\sum_{k=0}^{N-1}X[k]e^{j(2\pi/N)kn};n=0,1,2,\ldots,N-1\\
X[k]=\sum_{n=-0}^{N-1}x[n]e^{-j(2\pi/N)kn};k=0,1,2,\ldots,N-1
\end{eqnarray}

The sequence $X[k]$; $k=0,1,2,\ldots,N-1$ can be efficiently computed by a digital algorithm known as Fast Fourier Transform (FFT), from which the amplitude spectrum, $\vert X[k]\vert$ or ${\it abs}(X[k])$, and the phase spectrum, ${\it arg}(X[k])$ or ${\it angle}(X[k])$, are both obtained.

Next we obtain the DFT representation for each time series in the set by the 1024 points-FFT of the set matrix. Since each time series produced by the H\'enon mapping (with A $\in$ [0.2,1.4]) is a finite-duration bounded sequence, it satisfies the existence requirement for the DFT representation, that is,
\begin{eqnarray}
\sum_{n=-\infty}^{\infty}\vert x[n] \vert< \infty
\end{eqnarray}

We considered that is convenient, for some proposes, eliminating the dc component of signals produced by the H\'enon mapping, before the amplitude spectrum mesh plot was made.
\subsection{\bf Results}

Figures, from 1 to 3, depict a sample of three representative time series and their corresponding histograms. In order to give a better view of each time series it was only considered 128 data points of each discrete-time series for plotting but it was completely considered when its corresponding histogram was made. In either case, the particular value of $A$ parameter was specified.

The two-dimensional plot of the rectangular matrix produces a figure known as Feigenbaum or bifurcation diagram. It reflects the transition of a dynamical system towards chaos by a period-doubling process. This can be observed by running the program at appendix. Figure for bifurcation diagram is obtaining by plotting $A$ parameter values on the horizontal axis and sequence data values in the vertical axis. A threshold value $A\cong 1.0532$ is defined, beyong which the transition of the dynamical system to the chaotic behavior can be observed. The bifurcation diagram serves as a reference to control results, and it is usually applied in this research frame \cite{Aradi,Rodelsperger}.

Figure 5 depicts results for the largest Lyapounov exponent estimate $\lambda_1$, calculated for each of the time series in the set, as a function of the control parameter $A$. This reference is taken as a control for the discussion of the results obtained by applying the DFT representation, and it is too a particular aim of this work. 

Amplitude spectrum $\vert X[k]\vert$ or $abs(X[k])$ and phase spectrum $arg(X[k])$ or $angle(X[k])$, both for the discrete-time series matrix, may be plotted (see appendix). In either case, the {\it x} axis corresponds to the control parameter $A$, and the {\it y} axis corresponds to the frequency, from zero to the Nyquist frequency value. In order to obtain the best view of the DFT representation amplitude spectrum mesh plot, the dc component on the signal was eliminated. Figure 6 shows the phase spectrum mesh plot of the DFT representation of the discrete-time series set. 
\section{\bf Discussion}

The application of the DFT representation of a chaotic-type signal gives satisfactory results for determining the order-chaos transition in the H\'enon mapping dynamics. This conclusion is confirmed by comparing that phase spectrum mesh plot depicted in Figure 6 with that shown in Figure 5, where the behavior of the largest Lyapounov exponent $\lambda_1$ with the control parameter $A$ is shown. A critical value for control parameter $A$ can be defined, beyong which sequences produced by the H\'enon mapping exhibit a chaotic behavior, in general. This threshold value is close to 1.0532, as is obtained from Figure 5. From the Figure 5, it can deduce the value 0.618 bit/iter for LLE, which correponding to a value 1.4 for control parameter A. It is in agree with those values obtained by Wolf et al \cite{Wolf}, 0.603 bit/iter, and by Iasemidis et al \cite{Iasemidis}, 0.577 bit/iter. 

Some intermediate ordered region in the chaotic behavioral region underly, but some authors \cite{Wolf} refers them as a consequence of the computational system limitations rather than a real behavior of a dynamical system. However, this relationship between order and chaos can be observed from the dependence of the largest Lyapounov exponent $\lambda_1$ with control parameter, $A$, whenever the largest Lyapounov exponent value tends to zero at bifurcation points embedded in chaotic region. The behavior of a dynamical system, for  control parameter values correponding to largest Lyapounov exponent close to zero, is not chaotic. The bifurcation diagram was builded up several times, changing the initial condition, and changing the step value for the control parameter, and the feature of the bifurcation diagram remains. It can be remarked that the order-chaos transition for the value $A\cong 1.0532$ can be observed either in DFT representation amplitude spectrum mesh plot and in DFT representation phase spectrum mesh plot.

On the other hand, Figures from 1 to 3 show the time domain representation of some of time series and their corresponding histogram. From Figure 1($A=0.99$), it can be deduced the bifurcation effect from their histogram. Note that for Figures 2($A=1.17$) and 3($A=1.24$) the control parameter satisfies the condition $A>1.0531$ in either case, and that Figure 3 corresponds to a value for the largest Lyapounov exponent close to zero, for which the discrete-time series is not chaotic. It can be remarked that the bifurcation diagram analysis is used in studies of the behavior of a dynamical system that tends to chaos through period-doubling procedure \cite{Aradi}.

Applying DFT representation to a discrete-time series in a set is recommended for qualitative detecting the nature (ordered or chaotic) of the behavior of the system, and this can be added to the metric tools of the nonlinear dynamical analysis methods. Any method that allow us to evaluate the behavior of a dynamical system, and that can be applied to a relatively short discrete-time series, must be taken into account as a practical method.
\section{Conclusions}

The application of the discrete-time series DFT representation can be made over a chaotic-type discrete-time series, nevertheless this time series is to be nonlinear one. This method have neither no hard requirement about the length of the experimental data sequence. If one deals with a nonlinear dynamical system that can modify its behaviour between order and chaos, a discrete-time series produced by the system reflects this change, and the DFT representation exhibits in its amplitude spectrum and in its phase spectrum the change in the system too. Any method, that allow us to give some criteria about the behaviour of a dynamical system and that can operate over a relatively short discrete-time series, acquires practical importance. The DFT representation can be used for complementing the results obtained by applying the most powerfull methods of the nonlinear dynamical analysis.

\section{Appendix}
In order to see the bifurcation diagram, a finest version of all figures including the DFT representation phase spectrum mesh plot and the DFT representation amplitude mesh plot, the following program may be runned on MatLab.\\
(For running on MatLab, write a percent symbol at the begining of each comment line) 

Comment: To producing a family consisting of 120 time series, each of one having 1024 data points.\\
Comment: Figures for DFTHM paper\\
Comment: To producing a family consisting of 120 time series, each of one having 1024 data points.\\
clear\\
close all\\
Comment: Data\\
$alpha=0.2$; $del=0.01$; $nalpha=121$; $x=0.098$; $y=0.068$; $N=1024$; $beta=0.3$;\\
for $j=1:nalpha$;\\
   $Alpha(j)=alpha$;\\
   $alpha=alpha+del$;\\
end;\\
for $j=1:nalpha$;\\
   for $i=1:N$;\\
      $A(i,j)=x$; $z=x$;\\
      $B(i,j)=y$;
      $x=1-Alpha(j)*x^2+y$;\\
      $y=beta*z$;\\
   end;\\
end;\\
Comment:Figures selected\\
$j=[80\; 98\; 105\;]$;\\
for $k=1:3$;\\
   $alfa(k)=Alpha(j(k))$;\\
end;\\
disp('alpha values are:'); alfa,\\
disp('values of j equals to 80, 98, and 105 were selected');\\
Comment: $Alpha=[0.99\; 1.17\; 1.24]$;\\
pause(1); figure(1); subplot(211); plot(A(1:128,80));\\
title('Figure 1: Time series and its histogram, for $A=0.99$');\\
xlabel('Discrete-time index'); ylabel('Sequence values');\\
subplot(212); hist(A(1:128,80),40); xlabel('Sequence values');\\
ylabel('Relative frequency');\\
pause(1); figure(2); subplot(211); plot(A(1:128,98));\\
title('Figure 2: Time series and its histogram, for $A=1.17$');\\
xlabel('Discrete-time index'); ylabel('Sequence values');\\
subplot(212); hist(A(1:128,98),40); xlabel('Sequence values');\\
ylabel('Relative frequency');\\
pause(1); figure(3); subplot(211); plot(A(1:128,105));\\
title('Figure 3: Time series and its histogram, for $A=1.24$');\\
xlabel('Discrete-time index'); ylabel('Sequence values');\\
subplot(212); hist(A(1:128,105),40); xlabel('Sequence values');\\
ylabel('Relative frequency');\\
Comment:For optimal filling criterium\\
pause(1); figure(4);\\
title('Figure 4: Optimal filling of the phase space');\\
subplot(221); plot(A(1:N,114),B(1:N,114));\\
title('Figure 4a : $A=1.33$');\\
ylabel('First difference');\\
subplot(222); plot(A(1:N,116),B(1:N,116));\\
title('Figure 4b : $A=1.35$');\\
ylabel('First difference');\\
subplot(223); plot(A(1:N,118),B(1:N,118));\\
title('Figure 4c : $A=1.37$');\\
xlabel('Coordinate');\\
ylabel('First difference');\\
subplot(224); plot(A(1:N,120),B(1:N,120));\\
title('Figure 4d : $A=1.39$');\\
xlabel('Coordinate');\\
ylabel('First difference');\\
Comment:Part of the program to obtain the bifurcation diagram.\\
pause(1); figure(8);\\
for $i=1:N$; hold on; plot(Alpha,A(i,:)); end; grid on;\\
title('Figure 8: Bifurcation diagram'); xlabel('Control parameter');\\
ylabel('Sequence values');\\
Comment:Part of the program for largest Lyapounov exponent ploting.\\
$xr=0.2:0.01:0.2+120*.01$;\\
\begin{eqnarray}
yexl=[-0.309\; -0.286\; -0.201\; -0.219\; -0.217\; -0.199\; -0.180\; -0.162\; -0.152\; -0.135...\nonumber\\
-0.111\; -0.098\; -0.077\; -0.066\; -0.044\; -0.032\; -0.013\; 0.001\; -0.034\; -0.075...\nonumber\\
-0.113\; -0.153\; -0.196\; -0.243\; -0.294\; -0.352\; -0.419\; -0.499\; -0.591\; -0.789...\nonumber\\
-1.002\; -0.496\; -0.769\; -1.021\; -0.969\; -0.889\; -0.887\; -0.827\; -0.897\; -0.937...\nonumber\\
-0.903\; -0.854\; -0.838\; -0.842\; -0.865\; -0.862\; -0.824\; -0.995\; -0.924\; -1.013...\nonumber\\
-0.867\; -0.634\; -0.874\; -0.917\; -0.904\; -1.187\; -1.233\; -1.183\; -0.957\; -0.727...\nonumber\\
-0.602\; -0.499\; -0.419\; -0.353\; -0.295\; -0.243\; -0.196\; -0.153\; -0.113\; -0.076...\nonumber\\
-0.041\; -0.007\; -0.043\; -0.132\; -0.241\; -0.402\; -0.642\; -1.661\; -0.711\; -0.395...\nonumber\\
-0.232\; -0.124\; -0.041\; -0.062\; -0.407\; -0.019\;\; 0.041\;\; 0.141\;\; 0.188\;\; 0.259...\nonumber\\
0.272\;\; 0.188\;\; 0.340\;\; 0.336\;\; 0.363\;\; 0.383\;\; 0.414\;\; 0.420\;\; 0.455\;\; 0.414...\nonumber\\
0.428\;\; 0.475\;\; 0.447\; -0.415\; -0.311\; -0.083\; -0.006\;\; 0.113\;\; 0.358\;\; 0.444...\nonumber\\
-0.248\;\; 0.294\;\; 0.453\;\; 0.513\;\; 0.499\;\; 0.526\;\; 0.501\;\; 0.502\;\; 0.549\;\; 0.570...\nonumber\\
0.618]\nonumber\\
\end{eqnarray}\\
pause(1); figure(5);\\ 
plot(xr,yexl,'kx'); hold on; plot(xr,yexl,'k:'); grid on;\\
xlabel('Control parameter'); ylabel('Largest Lyapounov exponent');\\
title('Figure 5: Largest Lyapounov exponent behaviour with A');\\
legend('Experimental value','Behaviour',0); zoom on;\\
Comment:Part of the program for calculating and ploting the amplitude spectrum mesh and the phase spectrum mesh of the DFT representation.\\
$A=A-ones(N,1)*mean(A)$; $k=9$; $f=linspace(0,pi,N/8)$;\\
for $k=0:N/8-1$; $m=2*k+1$; $Ai(k+1,:)=A(m,:)$; end;\\
$Ai=Ai-ones(N/8,1)*mean(Ai)$; $Alpha=linspace(0.2,1.40,(nalpha-1)/4)$;\\
for $j=0:(nalpha-1)/4$; $p=4*j+1$; $Aii(:,j+1)=Ai(:,p)$; end;\\
$Aii=Aii-ones(N/8,1)*mean(Aii)$; $ftAii=fft(Aii)$;\\
$amftAii=abs(ftAii(:,:))$; $anftAii=angle(ftAii(:,:))$;\\
pause(1); figure(7); mesh(Alpha,f,amftAii(1:128,1:30));\\
title('Figure 7: Amplitude spectrum mesh');\\
xlabel('Control parameter'); ylabel('Frequency');\\ 
pause(1); figure(6); mesh(Alpha,f,anftAii(1:128,1:30));\\
title('Figure 6: Phase spectrum mesh');\\
xlabel('Control parameter'); ylabel('Frequency');\\
view(-33,50);\\

\newpage
\begin{figure}[b]
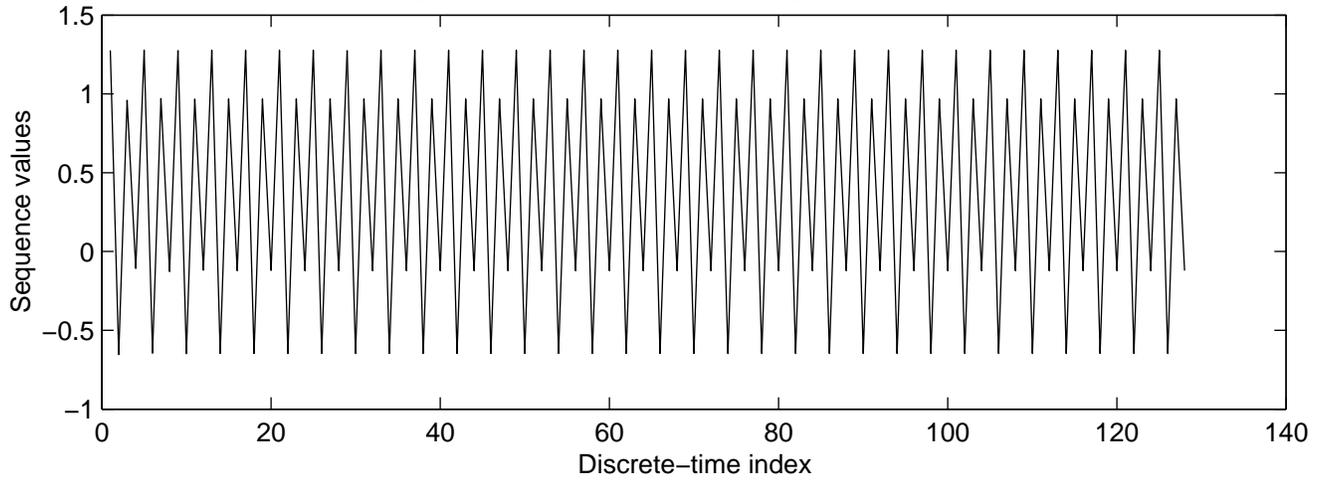

\caption{Time series and its histogram, for A=0.99}
\end{figure}
\begin{figure}[b]
\caption{Time series and its histogram, for A=1.17}
\end{figure}
\begin{figure}[b]
\caption{Time series and its histogram, for A=1.24}
\end{figure}
\begin{figure}[b]
\caption{Optimal filling of the phase space criterium}
\end{figure}
\begin{figure}[b]
\caption{Behavior of the largest Lyapounov exponent}
\end{figure}
\begin{figure}[b]
\caption{Phase spectrum mesh}
\end{figure}

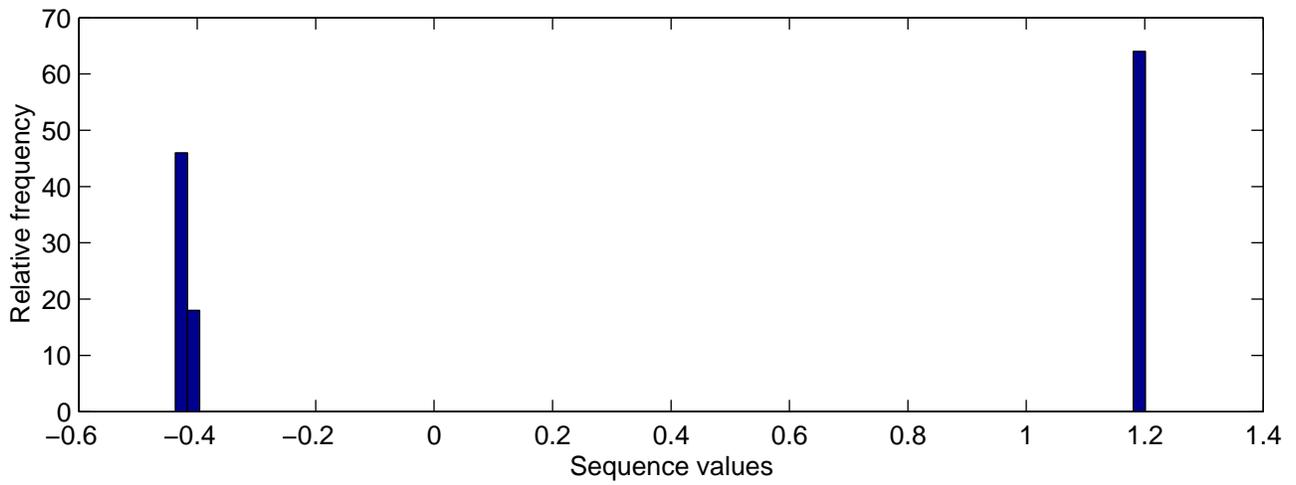
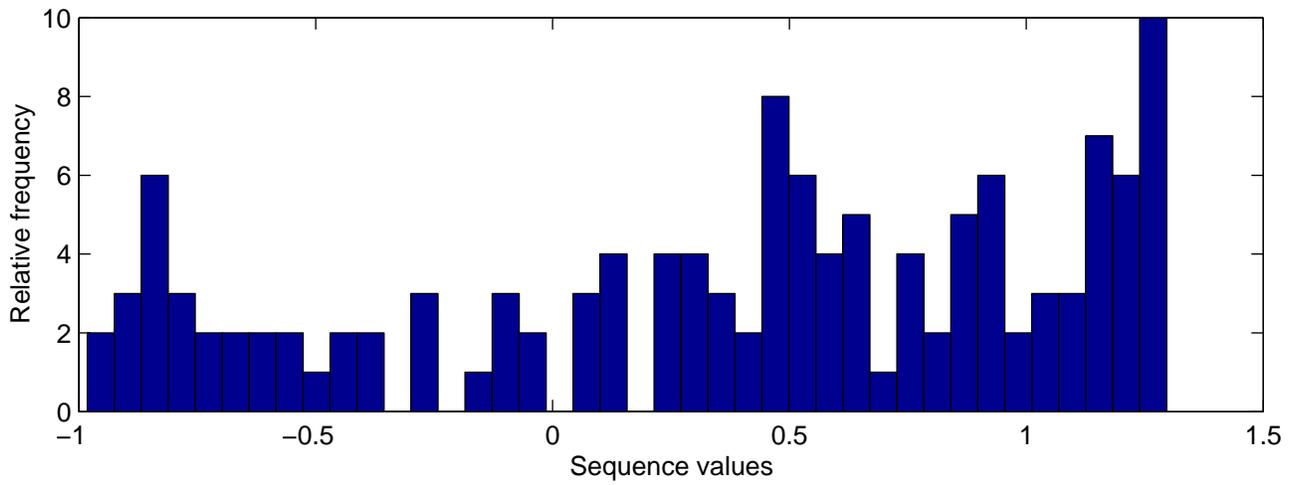
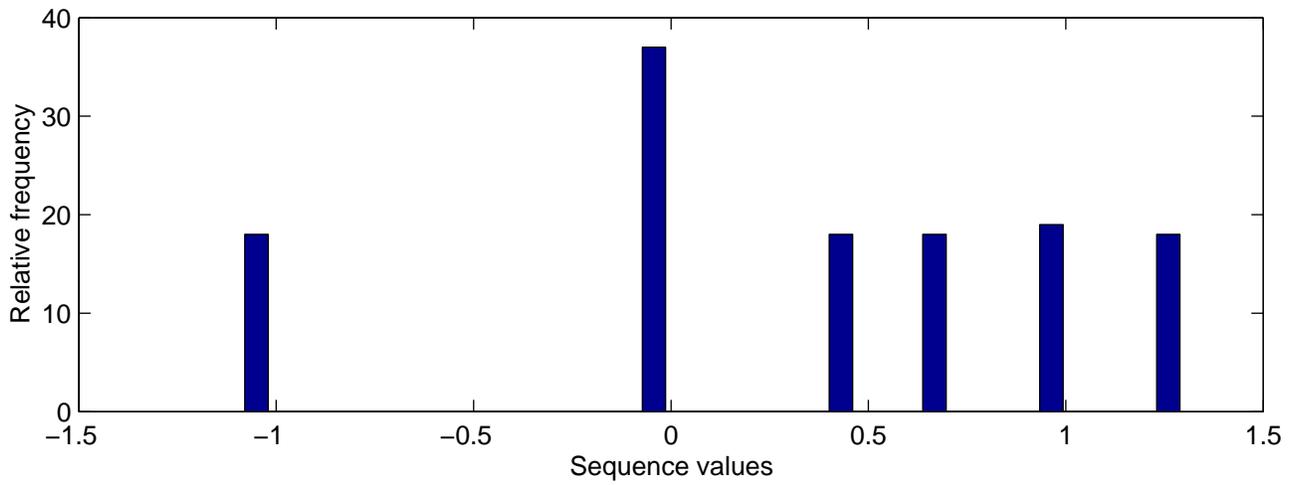
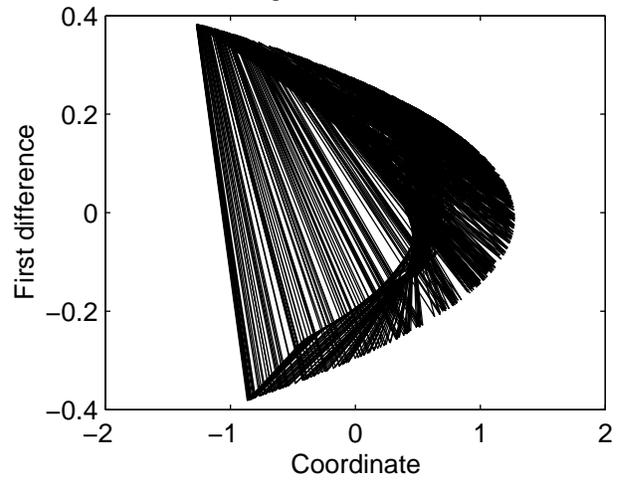
\end{document}